\documentclass[preprint, showpacs, aps, draft]{revtex4}
\usepackage{bm}

\begin{document}
\title{Information geometry, dynamics and discrete quantum mechanics\let\thefootnote\relax\footnotetext{Presented at MaxEnt 2012, the 32nd International Workshop on Bayesian Inference and Maximum Entropy Methods in Science and Engineering, July 15-20, 2012, Garching near Munich, Germany.}}

\author{Marcel Reginatto}
\affiliation{Physikalisch-Technische Bundesanstalt, Bundesallee 100,
38116 Braunschweig, Germany}

\author{Michael J. W. Hall}
\affiliation{Centre for Quantum Dynamics, Griffith University, Brisbane, QLD 4111, Australia\\}

\begin{abstract}
We consider a system with a discrete configuration space. We show that the geometrical structures associated with such a system provide the tools necessary for a reconstruction of discrete quantum mechanics once dynamics is brought into the picture. We do this in three steps. Our starting point is \emph{information geometry}, the natural geometry of the space of probability distributions. Dynamics requires additional structure. To evolve the $P^k$, we introduce coordinates $S^k$ canonically conjugate to the $P^k$ and a \emph{symplectic structure}. We then seek to extend the metric structure of information geometry, to define a geometry over the full space of the $P^k$ and $S^k$. Consistency between the metric tensor and the symplectic form forces us to introduce a \emph{K\"{a}hler geometry}. The construction has notable features. A complex structure is obtained in a natural way. The canonical coordinates $\psi^k = \sqrt{P^k} e^{iS^k}$ of the K\"{a}hler space are precisely the wave functions of quantum mechanics. The full group of unitary transformations is obtained. Finally, one may associate a Hilbert space with the K\"{a}hler space, which leads to the standard version of quantum theory. We also show that the metric that we derive here using purely geometrical arguments is precisely the one that leads to Wootters' expression for the statistical distance for quantum systems.
\end{abstract}


\maketitle

\section{Introduction}

In this paper, we consider systems with a discrete configuration space. In the presence of uncertainty, the state of a classical system will be described by a probability $P=(P^1,...,P^n)$, where $n$ is the number of available states. An analysis complementing the one carried out in Ref. \cite{RH11}, for continuous systems, leads to the remarkable result that the geometrical structures associated with such a system provide the tools necessary for a reconstruction of discrete quantum mechanics once dynamics is brought into the picture. The reformulation of Ref. \cite{RH11} presented here is nontrivial: it requires new geometrical insights because assumptions that are natural in the infinite dimensional case (e.g., ``spatial locality,'' the action of the Galilean group, etc.) are no longer available in the discrete case.

This reconstruction of discrete quantum mechanics has notable features. We show that the natural geometry of the space of probabilities in motion (i.e., taking dynamics into consideration) is a K\"{a}hler geometry. The canonical coordinates $\psi^k = \sqrt{P^k} e^{iS^k}$ of the K\"{a}hler space are precisely the wave functions of quantum mechanics. The full group of unitary transformations is obtained. Finally, one may associate a Hilbert space with the K\"{a}hler space, which leads to the standard version of quantum theory. We note that the K\"{a}hler space metric that we derive here using purely geometrical arguments is precisely the one that leads to Wootters' expression for the statistical distance for quantum systems \cite{W81}, which he derived using a completely different argument based on the concept of distinguishability. Comparison of our approach to papers of Mehrafarin \cite{M05} and Goyal \cite{G08,G10}, which also start from classical information geometry, will be made elsewhere.

\section{Information geometry}\label{ptig}

The starting point is a classical system which has associated with it $n$ different states. The probability that the system is in state $i$ is given by $P^i$, $i=1,...,n$, with $P^i \geq 0$ and $\sum_i P^i=1$. There is a natural line element given by
\begin{equation}\label{naturalLineElement}
ds^2 = G_{ij} \; dP^i \; dP^j  = \frac{\alpha}{2P^i} \; \delta_{ij} \; dP^i \; dP^j
\end{equation}
where $\alpha$ is a constant. The value of this constant can not be determined {\it a priori}; it is usually set to $\frac{1}{2}$. We do not make this assumption here but instead allow $\alpha$ to be a free parameter. The metric $G_{ij}$ is known as the \emph{information metric},
\begin{equation}\label{metricP}
G_{ij} = \frac{\alpha}{2P^i} \; \delta_{ij}.
\end{equation}

The line element of Eq. (\ref{naturalLineElement}) leads to a concept of distance on a probability space. This distance seems to have been introduced into statistics by Bhattacharyya \cite{B43, B46} as a way of providing a measure of divergence for multinomial probabilities \cite{G90}. Wootters calls it the \emph{statistical distance} \cite{W81}. To derive the statistical distance from the metric $G_{ij}$, consider two points in probability space, $P_A$ and $P_B$, joined by a curve $P^i(t)$, $0 \leq t \leq 1$, and write the expression for the length $l$ of the curve in the form
\begin{equation}\label{SDG}
l = \int_0^1 dt \sqrt{G_{ij}\frac{dP^i(t)}{dt}\frac{dP^j(t)}{dt}} \; .
\end{equation}
The statistical distance is defined as the shortest distance between $P_A$ and $P_B$. To compute the statistical distance, it is convenient to do the change of coordinates $X^i=\sqrt{P^i \,}$. Then
\begin{equation}
l =  \sqrt{2 \alpha}  \; \int_0^1 dt \sqrt{\sum_{i=1}^n \left[\frac{dX^i(t)}{dt}\right]^2 \,} \; .
\end{equation}
Since the curve $P(t)$ is assumed to lie in the probability space, it must satisfy the condition $\sum_{i=1}^n P^i(t) = \sum_{i=1}^n [X^i(t)]^2 = 1$; that is, the curve must lie on a unit n-dimensional sphere in the $X$ space. The shortest distance on the n-dimensional sphere is equal to the angle between the unit vectors $X_A$ and $X_B$. This leads immediately to
\begin{equation}
d(P_A,P_B) = \sqrt{2 \alpha} \, \cos^{-1} \left( \sum_{i=1}^n X_A^i \, X_B^i \right)= \sqrt{2 \alpha} \, \cos^{-1} \left( \sum_{i=1}^n \sqrt{P_A^i} \, \sqrt{P_B^i}  \right).
\end{equation}
which agrees with the expressions in the papers of Bhattacharyya and Wootters provided $\alpha$ is set to the standard value of $\frac{1}{2}$.

The statistical distance does not play a fundamental role in our discussion. Nevertheless, we make reference to it here because it clarifies the relation of our work to that of Wootters.

\section{Dynamics, symplectic geometry, and observables}

We now set the probabilities in motion. We assume that the dynamics of $P^i$ are generated by an action principle and we introduce additional coordinates $S^i$ which are canonically conjugate to the $P^i$ and a corresponding Poisson bracket for any two functions $F(P,S)$ and $G(P,S)$,
\begin{equation}\label{PoissonBrackets}
\left\{ F,G\right\} = \sum_i \left(
\frac{\partial F}{\partial P^i} \frac{\partial G}{\partial S^i}
 - \frac{\partial F}{\partial S^i}  \frac{\partial G}{\partial P^i} \right).
\end{equation}
As is well known, the Poisson bracket can be rewritten geometrically as
\begin{equation}
\left\{ F,G\right\} = \left( \partial F/\partial P\, , \; \partial F/\partial S\right) \, \Omega \, \left(
\begin{array}{c}
\partial G/\partial P \\
\partial G/\partial S
\end{array}
\right) ,
\end{equation}\nonumber
where $\Omega$ is the corresponding symplectic form, given in this case by
\begin{equation} \label{Omega}
\Omega=\left(
\begin{array}{cc}
0 & \textbf{1} \\
-\textbf{1} & 0
\end{array}
\right),
\end{equation}
where \textbf{1} is the unit matrix in $n$ dimensions. We thus have a symplectic structure and a corresponding {\it symplectic geometry\/} (which is why this formulation of the dynamics is a natural one for a geometric approach). The equations of motion for $P^i$ and $S^i$ are given by $\dot{P^i} = \left\{ P^i,H\right\}$, $\dot{S^i} = \left\{ S^i, H \right\}$ where ${H}$ is the Hamiltonian that generates time translations.

The observables of the theory are functions $A(P,S)$ of the coordinates $P^i$ and $S^i$. Certain restrictions are imposed on them, so not every function is an observable. For example, the infinitesimal canonical transformation generated by any observable $A$ must preserve the normalization and positivity of $P$. This implies the two conditions \cite{H04}
\begin{equation}
A(P,S+c) = A(P,S),~~~~~\partial A / \partial S^i = 0 ~ \textrm{if} ~
P^i=0.
\end{equation}
Note that the first condition implies gauge invariance of the theory under $S^i \rightarrow S^i + \chi$, where $\chi$ is a constant \cite{H04}.

\section{K\"{a}hler geometry}

We now want to consider the following question: Can we extend the metric $G_{ij}$ in Eq.~(\ref{metricP}), which is only defined on the $n$-dimensional subspace of probabilities $P^i$, to the full $2n$-dimensional phase space of the $P^i$ and $S^i$? It can be done, but certain conditions which ensure the compatibility of the metric and symplectic structures have to be satisfied. These conditions are equivalent to requiring that the space have a K\"{a}hler structure (see the Appendix of Ref. \cite{RH11} for a proof). We are led then to the beautiful result that the natural geometry of the space of probabilities in motion is a {\it K\"{a}hler geometry\/}.

A K\"{a}hler structure brings together metric, symplectic and complex structures in a harmonious way. To define such a space, introduce a complex structure $J_{\ b}^{a}$ and impose the following conditions \cite{G82},
\begin{eqnarray}
\Omega _{ab} &=& g_{ac}J_{\ b}^{c} \;,  \label{c1}\\
J_{\ c}^{a}g_{ab}J_{\ d}^{b} &=& g_{cd} \;,  \label{c2}\\
J_{\ b}^{a}J_{\ c}^{b} &=& -\delta _{\ c}^{a} \;.  \label{c3}
\end{eqnarray}
Eq. (\ref{c1}) is a compatibility equation between the symplectic structure $\Omega _{ab}$ and the metric $g_{ab}$, Eq. (\ref{c2}) is the condition that the metric should be Hermitian, and Eq. (\ref{c3}) is the condition that $J_{\ b}^{a}$ should be a complex structure.

We derive the solutions to these equations. The metric over the subspace of probabilities is the information metric, Eq. (\ref{metricP}). The metric over the full space will take the form
\begin{equation}
g_{ab}=\left(
\begin{array}{cc}
\textbf{G} & \textbf{E} \\
\textbf{E}^{T} & \textbf{F}
\end{array}
\right),
\end{equation}
where $\textbf{G}=\texttt{diag}(\frac{\alpha}{2P^i})$, and \textbf{E} and \textbf{F} are $n \times n$ matrices that need to be determined using the K\"{a}hler conditions and the expression for $\Omega_{ab}$, Eq. (\ref{Omega}). A matrix calculation leads to general forms for the metric $g_{ab}$ and the complex structure $J_{\ b}^{a}$. These are
\begin{equation}\label{gj_general}
g_{ab}=\left(
\begin{array}{cc}
\textbf{G} & \textbf{A}^{T} \\
\textbf{A} & (\textbf{1}+\textbf{A}^2)\textbf{G}^{-1}
\end{array}
\right),  ~~~~~
J_{\ b}^{a}=\left(
\begin{array}{cc}
\textbf{A} & (\textbf{1}+\textbf{A}^2)\textbf{G}^{-1} \\
-\textbf{G} & -\textbf{G}\textbf{A}\textbf{G}^{-1}
\end{array}
\right).
\end{equation}
where the $n \times n$ matrix \textbf{A} satisfies $\textbf{G}\textbf{A}\textbf{G}^{-1} =\textbf{A}^{T}$ but is otherwise arbitrary.

The K\"{a}hler conditions restrict the form of the metric $g_{ab}$ but leave the matrix $\textbf{A}$ (and therefore the geometry) undetermined. To fix the geometry, it is necessary to introduce an additional condition. As we show in the Appendix, there is a natural requirement based on the idea of invariance of the metric under the motions generated by the observables of the theory. This condition leads to $\textbf{A}=0$ and, as shown below, to a very simple geometry.  The condition $\textbf{A}=0$ alternatively follows by assuming the extended metric has zero curvature, analogously to the classical information metric $G_{ij}$.

\section{Complex coordinates}

We set $\textbf{A}=0$ and consider the K\"{a}hler structure given by
\begin{equation}\label{OgJPS}
\Omega _{ab}=\left(
\begin{array}{cc}
0 & 1 \\
-1 & 0
\end{array}
\right),  ~~~~~
g_{ab}=\left(
\begin{array}{cc}
\textbf{G} & 0\\
0 & \textbf{G}^{-1}
\end{array}
\right),  ~~~~~
J_{\ b}^{a}=\left(
\begin{array}{cc}
0 & \textbf{G}^{-1} \\
-\textbf{G} & 0
\end{array}
\right).  \label{SPogj}
\end{equation}
We now carry out the Madelung transformation, $\psi^i =\sqrt{P^i}\exp (iS^i/\alpha)$, $\bar{\psi}^i=\sqrt{P^i}\exp (-iS^i/\alpha)$. In terms of these complex coordinates, the tensors that define the K\"{a}hler geometry, Eqs. (\ref{SPogj}), take the standard form which is characteristic of a flat-space \cite{G82},
\begin{equation}\label{OgJpsipsistar}
\Omega _{ab}=\left(
\begin{array}{cc}
0 & i\alpha \textbf{1} \\
-i\alpha \textbf{1} & 0
\end{array}
\right),~~~~~
g_{ab}=\left(
\begin{array}{cc}
0 &  \alpha \textbf{1} \\
\alpha \textbf{1} & 0
\end{array}
\right),~~~~~
J_{\ b}^{a}=\left(
\begin{array}{cc}
-i \textbf{1} & 0 \\
0 & i \textbf{1}
\end{array}
\right).
\end{equation}
This shows that the simplest geometrical formulation of the space of probabilities in motion has a natural set of fundamental variables, $\psi^i$ and $\bar{\psi}^i$. If we set the constant $\alpha$ equal to $\hbar$, these fundamental variables are precisely the {\it wave functions\/} of quantum mechanics.

This is a remarkable result because we have not introduced {\it any} assumptions that concern quantum mechanics, only geometrical arguments.

\section{Statistical distance in the K\"{a}hler space}

We now want to introduce a new expression for statistical distance which will be valid in the K\"{a}hler space. In going from the $n$-dimensional space of probabilities $P^i$ to the full $2n$-dimensional phase space of the $P^i$ and $S^i$, the metric has been extended, and this should be taken in consideration when carrying out the generalization of the statistical distance.

Consider two points $\psi_A$ and $\psi_B$ representing states in this space which are joined by a curve $\psi^i(t)$, $0 \leq t \leq 1$. Combining Eq. (\ref{SDG}) with Eq. (\ref{OgJpsipsistar}), the correct generalization of the expression for the distance $l$ will be given by
\begin{equation}\label{GSDG}
l = \int_0^1 dt \sqrt{g_{ij}\frac{d\psi^i(t)}{dt}\frac{d\bar{\psi}^j(t)}{dt}} \;
= \sqrt{2 \alpha} \int_0^1 dt \left| \left(\frac{d\psi(t)}{dt} , \frac{d{\psi}(t)}{dt}\right) \right| \;.
\end{equation}
where we have introduced the notation $\left| \left(\psi , {\psi} \right) \right| = \sqrt{\sum_{i}\psi^i \bar{\psi}^i} $.

The statistical distance in the K\"{a}hler space is defined as the shortest distance computed with Eq. (\ref{GSDG}). Since the curve $P(t)$ is assumed to lie in the probability space, then $\psi(t)$ must satisfy the condition
\begin{equation}
1 = \sum_{i}\psi^i(t) \bar{\psi}^i(t),
\end{equation}
that is, the curve must lie on the unit sphere in the \{$\psi(t),\bar{\psi}(t)$\} space. The shortest distance on the unit sphere is equal to the angle between the unit vectors $\psi_A$ and $\psi_B$. This leads immediately to the expression for the statistical distance that appears at the end of section III of Wootters' paper (no equation number) provided $\alpha$ is set to the standard value of $\frac{1}{2}$,
\begin{equation}
d(\psi_A,\psi_B) = \sqrt{2 \alpha} \; cos^{-1} \left| \left(\psi_A , {\psi_B} \right) \right|.
\end{equation}

In this way we provide a very brief, geometrical derivation of Wootters' expression for the statistical distance in quantum mechanics, which he derived using a completely different argument based on distinguishability \cite{W81}.

\section{Group of linear unitary transformations}

We now examine the transformations which are allowed by the theory. As pointed out briefly in the discussion on observables, there are some basic conditions that must be satisfied; we list them and derive the group of transformations that is consistent with such conditions.

The first requirement is that the transformations preserve the normalization of the probability, $\sum_{i}P^i = \sum_{i}\psi^i \bar{\psi}^i = 1$.

The second requirement is that the metric be form invariant under those transformations; i.e., that the line element $d\sigma^2 = 2 \alpha \sum_j d\bar{\psi}^j d\psi^j$ of the K\"{a}hler space is preserved by the transformations.

Requiring normalization of the probability and metric invariance leads to the group of rotations on the $2n$-dimensional sphere. Such rotations are linear with respect to $\psi^j$ and $\bar{\psi}^j$. For an infinitesimal transformation, it follows that
\begin{equation}
\dot{\psi}^j =-i\frac{\partial{H}}{\partial \bar{\psi}^j},~~~~\dot{\bar{\psi}^j} =  i\frac{\partial{H}}{\partial \psi^j},
\end{equation}
are linear in $\psi$ and $\bar{\psi}$, where $H$ is the Hamiltonian that generates the motion. Then $H$ must be of the form
\begin{equation}\label{H2nR}
H = E(t) +  \sum_{j,k} \left[ M_{jk} \bar{\psi}^j \psi^k + N_{jk} \psi^j \psi^k + \bar{N}_{jk} \bar{\psi}^j \bar{\psi}^k \right]
\end{equation}
where $E(t)$ is a arbitrary function of time, $M$ is Hermitian, and $N$ is symmetric.

The third and final requirement is that we only consider rotations on the $2n$-dimensional sphere that are compatible with the equations of motion. As pointed out in the discussion on observables, conservation of probability requires that the ensemble Hamiltonian be invariant (up to an additive constant) under $S^j\rightarrow S^j+\chi$, since to first-order \cite{H04}
\begin{equation}
0=\epsilon \sum_j \dot{P}^j = \epsilon \sum_j \frac{\partial H}{\partial S^j} = H(P,S+\epsilon)-H(P,S).
\end{equation}
This condition, when written in terms of complex coordinates, is equivalent to invariance of the Hamiltonian under $\psi\rightarrow\psi e^{i\chi}$. Using the notation $Q:=\sum N_{jk} \psi^j\psi^k$, Eq. (\ref{H2nR}) leads to the equality
\begin{equation}
\left[ Q e^{2i\chi} +\bar{Q} e^{-2i\chi} \right] = 0
\end{equation}
which must be valid for all $\chi$. Differentiating with respect to $\chi$ gives the additional equality
\begin{equation}
2i\left[Q e^{2i\chi} -\bar{Q} e^{-2i\chi}\right] = 0.
\end{equation}
Combining these two expressions leads to $Q=0$.  Since this must hold for all $\psi$, it follows that $N_{jk}\equiv 0$, i.e., the ensemble Hamiltonian has the Hermitian form
\begin{equation}
H = E(t) +  \sum_{j,k}  M_{jk} \bar{\psi}^j \psi^k
\end{equation}
as desired.

This shows that the group of transformations of the theory is precisely the group of {\it linear unitary transformations}. Note that all that is used here, in moving from all rotations on the $2n$-dimensional sphere to the subset of unitary transformations, is (i) the conservation of probability and (ii) that the equations of motion follow from an action principle.

\section{Hilbert space formulation}

There is a standard construction that associates a complex Hilbert space with any K\"{a}hler space. Given two complex vectors $\phi^i$ and $\varphi^i$, define the Dirac product by \cite{K79}
\begin{eqnarray}
\langle \phi |\varphi \rangle &=&\frac{1}{2 }\sum_i \left\{ \left(
\phi^i,\bar{\phi}^i \right) \cdot \left[
g+i\Omega \right] \cdot \left(
\begin{array}{c}
\varphi^i \nonumber\\
\bar{\varphi}^i
\end{array}
\right) \right\}  \\
&=&\frac{1}{2 }\sum_i \left\{ \left( \phi^i,\bar{\phi}^i \right) \left[ \left(
\begin{array}{cc}
0 & \textbf{1} \\
\textbf{1} & 0
\end{array}
\right) +i\left(
\begin{array}{cc}
0 & i \textbf{1} \\
-i \textbf{1} & 0
\end{array}
\right) \right] \left(
\begin{array}{c}
\varphi^i \nonumber\\
\bar{\varphi}^i
\end{array}
\right) \right\}  \nonumber\\
&=& \sum_i \bar{\phi}^i \varphi^i
\end{eqnarray}

This suggests that the Hilbert space structure of quantum mechanics is perhaps not as fundamental as its geometrical structure.

\section{Concluding remarks}

We have shown that the Hilbert space formulation of discrete quantum theory emerges from the geometry of probabilities in motion. The basic elements that go into this geometrical reconstruction of discrete quantum mechanics are the natural metric on the space of probabilities (information geometry), the description of dynamics using a Hamiltonian formalism (symplectic geometry), and requirements of consistency (K\"{a}hler geometry).

We summarize some of the remarkable features of this construction. The wave functions of quantum mechanics, $\psi^k = \sqrt{P^k} e^{iS^k}$, appear as the natural complex coordinates of the K\"{a}hler space that describes the geometry of probabilities in motion. The full group of unitary transformations is derived based on consistency requirements. And, finally, a Hilbert space may be associated with the K\"{a}hler space of the theory, which leads to the standard version of quantum theory.

We have shown that we can derive Wootters' statistical distance for quantum mechanics, Eq. (\ref{GSDG}), in a purely geometrical way. We have commented in this brief paper on the connection between our work and that of Wootters; it would also be interesting to investigate the connection of our work to papers of Mehrafarin \cite{M05} and Goyal \cite{G08, G10} which also take an information-geometrical approach. One of the main differences seems to come from the emphasis that we place on finding a geometrical description for the space of probabilities {\it in motion}. The use of an action principle to describe the dynamics of the probabilities $P^i$ introduces geometrical structure that is quite powerful: we immediately get a doubling of the dimensionality of the space (i.e., $\{P^i\} \rightarrow \{P^i,S^i\}$) and, in addition, we end up with a complex structure and unitary transformations, all of them ingredients that are essential for quantum mechanics. In our formalism, these are mainly the result of consistency requirements which we impose to ensure the peaceful coexistence of information geometry and symplectic geometry.

It appears then that discrete quantum mechanics stands at the intersection of information geometry, symplectic geometry, and K\"{a}hler geometry.

\appendix

\section{The condition $\textbf{A}=0$}\label{AppTCAE0}\nonumber

To fix the matrix $\textbf{A}$ that appears in Eqs. (\ref{gj_general}) we need to supplement the K\"{a}hler conditions, Eqs. (\ref{c1}-\ref{c3}), with an additional condition. We make use of the fact that motion in the probability space (for example, the trajectory in phase space generated by an observable) preserves the normalization $\sum_i P^i=1$. To look more closely at this constraint, we introduce the new set of $2n$ real coordinates
\begin{equation}
x^i=\sqrt{2 \alpha P^i} \cos \left(S^i / \alpha\right),~~~~~
y^i=\sqrt{2 \alpha P^i} \sin \left(S^i / \alpha\right).
\end{equation}
One can check that this coordinate transformation is also a canonical transformation, $\{x^i,y^j\}=\delta^{ij}$, so the symplectic form remains invariant.

The constraint $\sum_i P^i=1$ takes the form $\sum_i \{(x^i)^2+(y^i)^2\}=2\alpha$ in the new coordinates. Therefore, motion is restricted to a $2n$-dimensional sphere in the space of the $x^i$ and $y^i$. We require that the metric be invariant under rotations (equivalently, invariant under the motions generated by observables); i.e., that it be spherically symmetric. In the coordinates $x^i$ and $y^i$, a spherically symmetric line element has the general form
\begin{equation}
d\rho^2 = f(r)\sum_i\left\{(dx^i)^2+(dy^i)^2\right\} + g(r)\left\{\sum_i(x^i dx^i+y^i dy^i)\right\}^2
\end{equation}
where $f$ and $g$ are arbitrary functions of $r=\sqrt{\sum_i\{(x^i)^2+(y^i)^2\}}$. In terms of the original coordinates $P^i$ and $S^i$, the spherically symmetric metric takes the form
\begin{equation}
d\rho^2 = f(\sum_i P^i) \sum_i\left\{\frac{\alpha}{2P^i}(dP^i)^2+\frac{2P^i}{\alpha}(dS^i)^2\right\} + g(\sum_i P^i) \left\{\alpha \sum_i dP^i \right\}^2
\end{equation}
The important point here is that the metric $d\rho^2$ does {\it not} have any mixed terms proportional to $dP^i dS^j$, and this means that we must set $\textbf{A}=0$ to satisfy our requirement of metric invariance.

\end{document}